\documentclass[amsmath,amssymb,nofootinbib,aps,prl,twocolumn,showpacs,superscriptaddress]{revtex4-1}

\usepackage{graphicx} 
\usepackage{bm}
\usepackage{url}
\usepackage{dcolumn}

\usepackage{mhchem}
\usepackage{wrapfig}
\usepackage{subfigure}
\usepackage{float}

\begin{document}

\title{Anti-jamming in a fungal transport network}

\author{Patrick C. Hickey}
\affiliation{NIPHT Ltd., 1 Summerhall, Edinburgh, EH9 1PL},
\author{Haoxuan Dou}
\author{Sierra Foshe}
\author{Marcus Roper} 
\affiliation{Depts. of Mathematics and Biomathematics, University of California, Los Angeles, CA 90095}

%

\date{\today}
\begin{abstract}
Congestion limits the efficiency of transport networks ranging from highways to the internet. Fungal hyphal networks are studied as an examples of optimal biological transport networks, but the scheduling and direction of traffic to avoid congestion has not been examined. We show here that the {\it Neurospora crassa} fungal network exhibits anticongestion: more densely packed nuclei flow faster along hyphal highways, and transported nuclei self-organize into fast flowing solitons. Concentrated transport by solitons may allow cells to cycle between growing and acting as transport conduits.
\end{abstract}
\maketitle
%
%
%
%

Congestion; the tendency for transport speeds to decrease as the density of traffic increases afflicts both highways and data networks \cite{jacobson1988congestion,*treiber2000congested}, and transport networks must be designed to prevent congestion-induced failure due to fluctuations in demand across the network \cite{chiu1989analysis,*yang1998models}. Maximizing transport efficiency for a given cost creates complex optimization problems \cite{banavar2000topology,*durand2006architecture}, and biological transport networks are thought to represent evolutionary solutions to these problems \cite{Tero:2010bx,*hu2013adaptation}. In particular, plant leaves and other natural networks may respond to fluctuations in demand by creating multiple redundant paths between the same nodes \cite{Corson:2010ee,*Katifori:2010uo}. \footnote{Although congestion decreases transport efficiency, it also provides a tool to design networks with some other optimality, such as uniform division of fluxes \cite{sheazebrafish}}. However, although transport efficiency of networks is influenced both by the geometry of the network, and by the rules that govern how traffic is routed through the network, the use of dynamic routing in biological networks has received less attention \cite{Alim:2013gn,*Hu2012bloodvessel} than growth and reconfiguration of the network geometry over longer time scales \cite{banavar2000topology,Corson:2010ee}. 

\begin{figure}
	\includegraphics[width=0.9\columnwidth]{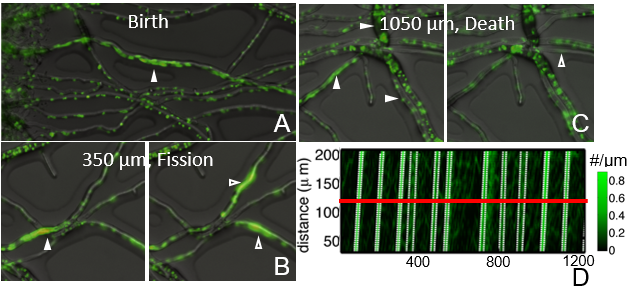}
	\caption{Nuclei transported within hyphal networks of {\it Neurospora crassa} spontaneously organize into rapidly moving solitons (see Supplementary Movie S1). (A) Solitons emerged spontaneously near the innoculation point. (B) At one hyphal branch point each soliton (filled arrow) split into two (hollow arrows). (C) For this train, solitons diassociated when three hyphae (filled arrows) merged into a single hypha with fast, dilute flow (hollow arrow). Scale bar: 50\,$\mu$m. (D) Heat-map of density (\#/length) of nuclei along a single hypha as a function of time and distance. Transport by solitons leads to oscillations in density when observed at a single point (e.g. along red line).}
\end{figure}

Filamentous fungi form branched and multiconnected networks of cells (called hyphae) that act as transport conduits for nutrients, fluid and organelles \cite{Lew05}. Hyphal networks adapt to minimize transport costs \cite{bebber2007biological,*heaton2010growth} and also to physically mix the cytoplasmic contents \cite{Roper:2013jr}. Typically in the model filamentous fungus {\it Neurospora crassa}, nuclei are uniformly dispersed through each hypha \cite{Roper:2013jr,Abadeh:2013dy}, but in narrow fast flowing hyphae, nuclei have been recently reported to travel together as dense groups \cite{Hickey2009comets} (Fig. 1). The causes of group formation have not previously been reported. In this Letter we show groups of nuclei are fast moving solitons that form due to anti-congestion within the hyphal network: nuclei travel faster in denser groups than they can individually. In contrast to traffic jams which grow by assimilating cars from the rear \cite{lighthill1955kinematic}, solitons are created by anti-jamming, that is by overtaking other nuclei. Nuclei continuously transition between transported and stationary phases; anti-congestion results from cooperation between nuclei that suppresses their transitions out of the transported phase. Scheduling of nuclear traffic into intermittent solitons may represent an evolutionary solution to a common problem of biological transport networks; that the links of the network must be remodeled over time without disrupting their function as transport conduits.

Following \cite{Hickey2009comets} we visualized the flow of nuclei in young ($<$ 1cm diameter) {\it hH1::gfp} {\it N. crassa} mycelia \cite{freitag2004gfp}, grown on solid Vogels minimal medium. This genetically altered strain expresses a GFP (green fluorescent protein)-tagged histone, making nuclei fluorescent. We followed dense groups of nuclei from their first emergence near the center of network through 1 mm or more of hyphae (Fig. 1 and movie S1). In agreement with \cite{Hickey2009comets} we found that dense groups traveled together across multiple hyphal links and could split into two separate groups where hypha branched (Fig. 1B). Nuclear groups ultimately dispersed only after traveling 1 mm or more through the network (Fig. 1C). 

We used a hybrid particle imaging velocimetry / particle tracking method \cite{kaitlynpiv} to track nuclear movements within and outside of dense groups. Nuclei traveling in groups were highly coherent, with no detectable difference in velocity between the front of the group and its rear (Fig 1D). Nuclear distributions were stable between different groups (Fig. 2A). Taken together, these features suggest that groups emerge represent stable traveling wave (soliton) solutions of the equations governing nuclear transport. 

We compared speeds of individual and anti-jammed nuclei for over 3000 nuclei (11 individual anti-jams) in a single 400\,$\mu$m stretch of hypha: In general speed increased with the number density of nuclei (Fig. 2B), and groups travel faster than individual nuclei; we therefore call them {\it anti-jams}. Thus, {\it N. crassa} transport networks perform oppositely to vehicle and data networks \cite{jacobson1988congestion,*treiber2000congested}, in which speeds decrease congestively as traffic density increases. We measured the velocities of individual nuclei that were at least 5\,$\mu$m apart from other nuclei: these nuclei could be divided by Otsu's method \cite{otsu1975threshold} into two populations by their velocity {\it stationary} ($v<5.8\,\mu$ms$^{-1}$) or {\it transported} ($5.8<v\lesssim 20\,\mu$ms$^{-1}$) (Fig. 2C). Velocities of nuclei propelled by bulk cytoplasmic flow are known to be nearly uniform across the entire hyphal cross-section \cite{Roper:2013jr,Abadeh:2013dy}; likely because cytoplasm flows like a polymer melt with a narrow slip layer along the cell membrane \cite{roper2015life}. Stationary nuclei were located at the periphery of the cell, were deformed by long tubular tethers and occasionally traveled backwards (Fig. 2C and inset) suggestive of active motor-driven positioning \cite{roux2002minimal,*mourino2006microtubule}.


Nuclei do not persist in their transported or stationary phases but freely transitioned between the two: on average nuclei remain attached to the cell membrane for 20s, and flowed with the cytoplasm was 7.8s between transitions (Fig. 2D). Attachment and detachment occurred at constant rates along the the hypha, suggesting that potential attachment sites are uniformly distributed. 
\begin{figure}
	\includegraphics[width=0.85\columnwidth]{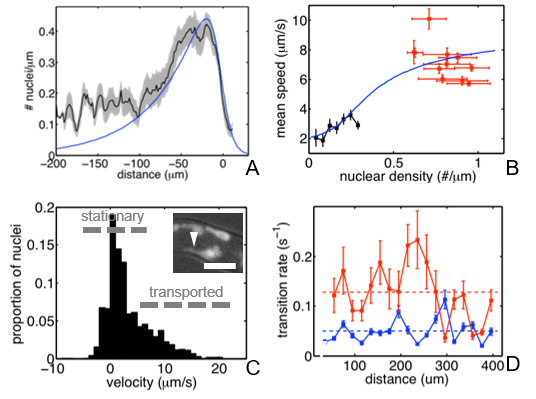}
	\caption{Solitons are produced by anti-jamming. (A) Similarity of nuclear distributions along the length of different solitons. The black curve gives the mean from 11 different solitons, measured at the same point in the hypha, and the gray shading shows the standard error. Blue curve shows an asymptotically calculated density profile. (B) The mean velocity of nuclear transport increases as traffic becomes denser. Black: nuclei not in anti-jams, red: anti-jams, blue: predicted velocity-density relationship from Eq. (\ref{eq:pde}) with no free parameters. (C) Nuclei can be divided into transported and stationary populations. Inset: Slow moving nuclei are located near the cell membrane and are deformed, suggestive of active transport by motor proteins. Scale bar: 5\,$\mu$m. (D) Nuclei freely transition between stationary and transported phases, and transition rates (red: $T\to S$, blue: $S\to T$) are uniform along the length of the hypha.}
\end{figure}

To explain flow anti-congestion we model the density of transported, $T$ and stationary, $S$, nuclei (measured in \#/length of hyphae). Although we do not observe attachment sites directly, we assume that a nucleus enters the stationary phase by attaching to an available site (density $b$) on the cortical microtubule network. Thus our starting model (described in the Electronic Supplementary Materials); models the mass action laws for a reversible reaction system:
\ce{$T+b$ <=>[k_{TS}][k_{ST}] $S$}
where $k_{ST}$, $k_{TS}$ are the rates of nuclear attachment and detachment respectively. We assume that transported nuclei travel with the cytoplasm, at velocity $U$, and stationary nuclei and attachment sites remain fixed on the cell membrane. We can derive analytic results if the length scale on which $S$ and $T$ vary exceeds $U/k_{ST}$; the travel distance lost by a nucleus that remains attached over the characteristic time $1/k_{ST}$. In this limit stationary and transported nuclei maintain a local equilibrium: $T =\frac{k_{ST}S}{k_{TS}(b^*-S)}$, where $b^*$ is the (constant) density of potential attachment sites (occupied and unoccupied, so $b+S = b^*$). The total density of nuclei in both phases $n=T+S$ obeys a flux conservation equation: $\frac{\partial n}{\partial t} + \frac{\partial}{\partial x}\left(n u(n)\right) = 0$, where $u(n)=UT/n$ is the mean speed of transport of a mixed population of nuclei. Physically, our model predicts anti-congestion because there are finitely many available attachment sites on the cell membrane; so if density of nuclei increases then an increasing fraction are in the transported phase.

However the PDEs for $S$ and $T$ predictive dispersive, rather than free-running anti-jams (see Electronic Supporting Materials). Indeed the local equilibrium equations can be transformed to Burgers' equation by change of variables: $\rho\equiv \frac{d}{dn}\left(u(n)n\right)$. Burgers' equation has no traveling wave solutions with the same left and right boundary conditions \cite{evans1998partial}. Specifically, since $u(n)$ increases with $n$, negative density gradients steepen with time -- denser groups of nuclei travel faster than the nuclei ahead of them, becoming denser and faster. But small positive density gradients flatten over time; the sparse nuclei at the rear of an anti-jam travel slower than the denser nuclei ahead of them, so the anti-jam disperses. 


 How do nuclei in anti-jams travel coherently? We wondered whether nuclei were adhering together \cite{minke1999microscopic}, so we dissected nuclear movements using high-speed (8.3 fps) confocal microscopy. Nuclei inside anti-jams rearranged continuously and fluidly (Fig 3A). We resolved the velocity fluctuations of pairs of almost touching (center-to-center distance $<2.5\,\mu$m) nuclei into tangential and normal components. If nuclei were adhering we expected the tangential velocity component to be larger because it is easier for pairs of nuclei to rotate or slip past each other than to separate, but tangential and normal velocities were indistinguishable (Fig. 3B).

Burgers' equation allows traveling wave solutions with different left and right boundary conditions \cite{evans1998partial}, i.e. if one of our fields; $b$, $S$ or $T$ is different ahead of and behind the anti-jam. There is no detectable difference in nuclear density on either side of an anti-jam so we hypothesized that attachment sites might be altered. Although we can not directly visualize these attachment sites, we can detect their effect on nuclear transport, by measuring velocities of nuclei in the wake of each anti-jam (Fig. 3C). Nuclei traveled faster on average immediately behind an anti-jam, dropping back to the average velocity for dilute nuclei only after a time $\tau = 20$s (Fig. 3C). Increases in nuclear velocity correspond to fewer stationary nuclei in the wake of the anti-jam. However, since $\tau \gg 7.8s$ (the average time for a nucleus to transition back to a stationary state), these data also require that nuclei transition more slowly to stationary state behind the anti-jam.

To explain the speed up data we assume that attachment sites are remodeled, or even lost (and then regenerated) following nuclear attachment. Thus, after a nucleus detaches from the cell periphery, an attachment site is not immediately available to accept another nucleus. We must therefore consider three attachment site densities: $b$, available, $S$, attached (equal to the number of attached nuclei), and $u$, unattached but unable to accept nuclei. We assume that potential attachment sites are uniformly distributed along the hypha, with density $b+S+u=b^*$. When written in dimensionless form (see Electronic Supplementary Materials) densities of nuclei and attachment sites are describe by a set of equations:
\begin{eqnarray} 
\frac{\partial T}{\partial t} + \frac{\partial T}{\partial x} &=& -\alpha_1 T b + S \nonumber \\
\frac{\partial S}{\partial t}&=& \alpha_1 Tb - S \label{eq:pde} \\
\frac{\partial b}{\partial t} &=& \alpha_2 (1-b-S) - \alpha_1 T b \nonumber
\end{eqnarray}
Our equations include two dimensionless rate constants: (i) $\alpha_1 = \frac{k_{TS} b^*}{k_{TS}}$; the ratio of the nuclear transition rates and (ii) $\alpha_2 = (k_{ST}\tau)^{-1}$; the ratio of the time a nucleus spends attached to the time before an attachment site can accept more nuclei. In the limit $\alpha_2\to \infty$ we obtain the limit of the equations analyzed previously, with $b+S=1$.
\begin{figure}
	\includegraphics[width=0.8\columnwidth]{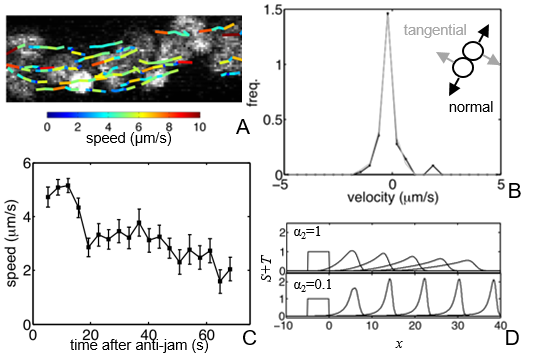}
	\caption{Stability of anti-jams requires that attachments sites remodel following nuclear detachment. (A) Nuclei within an anti-jam undergo continuous fluid rearrangements as the anti-jam moves through a hypha. Shown: nuclear trajectories over 0.6s, color-coded by velocity. (B) We resolved the velocities of pairs of touching nuclei into tangential (gray) and normal (black) components: the distributions of the two components are identical, suggesting no cohesion. (C) Nuclei travel faster for approximately $\tau\approx 20$s following the passage of a anti-jam. (D) A mathematical model that includes remodeling predicts solitons for sufficient small values of $\alpha_2=1/k_b\tau$.}
\end{figure}

Finite-difference numerical solutions of Eq.(\ref{eq:pde}) (see ESM) evolve to solitons for sufficiently small values of $\alpha_2$ but dispersed for larger values of $\alpha_2$ (Fig. 3D). Since small values of $\alpha_2$ seem to be necessary for traveling waves to be generated, we analyzed Eq.(\ref{eq:pde}) asymptotically in the $\alpha_2\to 0$ limit.  In this limit, attachment sites are monotonically depleted as the anti-jam completely passes. Nuclei therefore cooperate to eliminate all attachment sites, allowing anti-jams to travel through the network without dispersing. Soliton solutions exist, and the nuclear distribution within a soliton can be calculated analytically. For soliton solutions of Eq.(\ref{eq:pde}) with speed $c$, we may combine the PDEs for $T$ and $S$ to give: $(1-c)T-cS = 0$. We can solve for the distribution of nuclei within the anti-jam by using $b$ (the monotonically decreasing density of available attachment sites) as a dependent variable, and integrating:
\begin{equation}
S = \frac{1}{\alpha_1} \log b+\frac{c}{(1-c)}\left(1-b\right)
\end{equation}
These equation possesses a family of soliton solutions which can be parameterized either by the total number of nuclei or by the dimensionless velocity $0<c<1$. In the wake of the anti-jam the predicted nuclear density becomes exponentially small and the population $b$ returns to its far upstream value (see ESM).

Although our simulations and analytic results show that Eq.(\ref{eq:pde}) supports solitons, they are silent about whether solitons emerge spontaneously or must be created by modulating the distribution of nuclei. We analyzed the stability of uniform flow $S=S_0$, $T=T_0$, $b=b_0$ to infinitesimal disturbances of form: $S = S_0 + \tilde{S}e^{i(kx-\omega t)}$ for some wavenumber $k$ and frequency $\omega$, with similar representations for $T$ and $b$. Linearizing Eqn.(\ref{eq:pde}), we find that uniform flow is unstable to at most a finite range of $k$-values, and that at large values of $\alpha_2$ becomes absolutely stable (see ESM). Nonlinear simulations then showed that linearly unstable disturbances could grow into anti-jams (see ESM). 

Since $\alpha_1$ depends on the rate of nuclear/attachment site encounters, it is proportion to the cytoplasmic velocity. The predicted stability boundary when nuclear density $n_0=S_0+T_0$ and $\alpha_1$ were both varied separated hyphae carrying anti-jams and hyphae that did not (Fig. 4B). We further probed the role of linear stability in anti-jam formation and eventual dispersal by examining flows into and out of network junctions in which anti-jams occurred only on one side of the junction: each junction approximately crossed the predicted stability boundary (Fig. 4C-E, movies S3 and S4). To further confirm that the parameters of our model drive anti-jam formation we redirected flows by damaging hyphae with deionized water. Rerouting of flow from a large diameter, slow hypha, to a small diameter fast hypha, generated anti-jams, as predicted by the linear stability analysis (Fig. 4E). 

\begin{figure}
		\includegraphics[width=0.8\columnwidth]{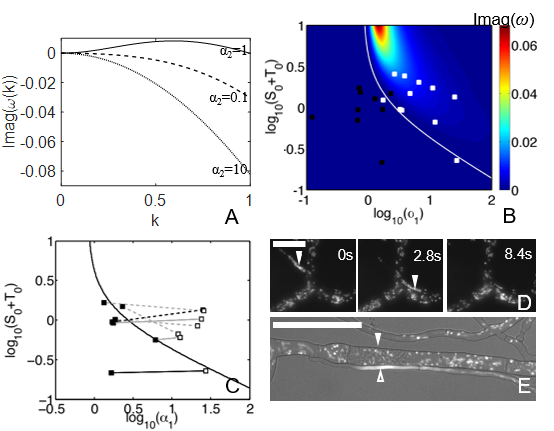}
	\caption{Nuclear flow spontaneously self-organizes into anti-jams.  (A) At any value of $\alpha_1$, there is at most a finite range of wavenumbers for which disturbances grow. Different curves correspond to different values of $\alpha_1$. (B) We performed a parameter-sweep over hypha with different traffic densities $S_0+T_0$ and velocities, $\alpha_1$, to determine which had unstable modes, and could therefore support anti-jams. The stability boundary (white line) divides anti-jam carrying hyphae (white squares), and hyphae without anti-jams (black squares). (C) Anti-jams form and die when flow pass between through junctions between hyphae on either side of the stability boundary. Solid lines show junctions across which anti-jams were created, and dashed lines junctions across which anti-jams dispersed. (D) Dispersal of anti-jam at the junction shown by the dashed black line in (C). Scale bar: 20 $\mu$m. (D) Nuclear flow redirected around a damaged trunk hypha (white arrow) into a fast narrow hypha (hollow arrow) spontaneously forms anti-jams (solid black line in (C)). Scale bar: 50 $\mu$m}
\end{figure}


In summary, nuclei within high traffic links of the fungal network self-organize into stationary and anti-jammed phases. Indeed since nuclei are likely exposed to the same proteins from their shared cytoplasm \cite{gladfelter2006asynchronous,roper2015life}, self-organization may provide a more robust mechanism for directing nuclear behaviors than signaling between nuclei.

Engineering of transport on the hyphal network to produce anti-congestion may provide two adaptive benefits: (1) A gain function for transport, that allows nuclei to be delivered faster in response to increases in demand, reducing the amount of redundant network links needed \cite{Corson:2010ee}. 
%
(2) Links in the hyphal network must grow over time to minimize transport costs \cite{bebber2007biological}. But, similar to repair work on an open freeway, it is likely that the continuous traffic of nuclei and other organelles will disrupt this growth. Anti-jams organize nuclear transport into very short pulses separated by long ($\sim$ 100s) intervals in which most ($\approx 85\%$) nuclei are retained at the cell periphery and stationary.  The intervals between anti-jams may allow growth and other functions to be performed, controlled by stationary nuclei. In particular, given protein translation rates of 8 amino acids/s \cite{karpinets2006rna}, and the close agreement of translation and transcription times \cite{milo2015cell}, the residence time of a nucleus at the cell membrane ($\sim 20s$) agrees well with the time needed to transcribe the key cell wall proteins RHO1 and RHO2 (195 and 200 amino acids respectively \cite{richthammer2012rho1}).






This work was supported by the Alfred P. Sloan Foundation and NSF grant DMS-1351860. We thank Inwon Kim, Louise Glass' group and Nick Read for discussions.

\end{document}


\hyphenation{ra-tion-aliz-ation}
\renewcommand*\thefigure{S\arabic{figure}}
\renewcommand*\theequation{S\arabic{equation}}
\renewcommand*\thetable{S\arabic{table}}

\newcommand\std{\rm std}
\newcommand\be{\boldsymbol{e}}
\newcommand\bx{\boldsymbol{x}}
\newcommand\bu{\boldsymbol{u}}
\newcommand\p{\partial}
\def\tp{\tilde{p}}
\newcommand\bzero{\boldsymbol{0}}
\newcommand\bv{\boldsymbol{v}}

\title{Supplementary Information for: Anti-jamming in a fungal transport network}
\author{Patrick C. Hickey}
\affiliation{NIPHT Ltd., 1 Summerhall, Edinburgh, EH9 1PL, United Kingdom}
\author{Haoxuan Dou}
\author{Sierra Foshe}
\author{Marcus Roper\footnote{To whom correspondence should be addressed. E-mail: <mroper@math.ucla.edu>}}
\affiliation{Depts. of Mathematics and Biomathematics, University of California, Los Angeles, CA 90095}

\date{\today}

\begin{abstract} This document contains the following Electronic Supplementary Information associated with the mathematical modeling described in the paper ``Anti-jamming in a fungal transport network'',
\begin{itemize}
\item[\ref{SSPDEanalysis}] Non-dimensionalization and existence of soliton solutions to equations for nuclear flow
\item[\ref{sec:anti-jams}] Asymptotic calculation of the soliton profile
\item[\ref{sec:stability}] Stability theory and numerical simulations of anti-jam formation from unstable uniform flow
\end{itemize}
\end{abstract}

\maketitle




\section{Non-dimensionalization and existence of soliton solutions in equations for nuclear flow \label{SSPDEanalysis}}

\subsection{Without remodeling of attachment sites \label{sec:noremodel}}

We start with rate equations for the transitions of nuclei between transported and stationary states, modeling potential attachment sites as an exhaustible resource. We assume that flowing nuclei travel at mean velocity $U$ (the velocity of cytoplasmic transport) \footnote{Previous measurements \cite{Abadeh:2013dy,Roper:2013jr}, have shown that there is considerable variance in velocities between flowing nuclei, due mainly to collisions with other organelles such as vacuoles within the crowded cytoplasm. However, there is no systematic variation in velocities across the hyphal cross-section, and in particular no evidence of systematic variation in velocity between the center of the hyphae and the cytoplasm at the cell wall, i.e. no Poiseuille velocity profile \cite{Batc67}. Fluctuations in velocity due to organelle-organelle collisions can be averaged over to obtain a single, average speed, governing all nuclear transport in each hypha.}. We write $T$, $b$ and $S$ respectively for the density (in $\#/\mu$m length of hypha) of flowing nuclei, available attachment sites, and dimers formed by nuclei and attachment sites. If the rate constants for attachment and detachment are $k_{TS}$ and $k_{ST}$ respectively, then the three populations are governed by a set of PDEs:
\begin{eqnarray} 
\frac{\partial T}{\partial t} + U\frac{\partial T}{\partial x} &=& -k_{TS} T b + k_{ST} S \nonumber \\
\frac{\partial S}{\partial t}&=& k_{TS} Tb - S \label{eq:pdediml} \\
\frac{\partial b}{\partial t} &=& k_{ST} S - k_{TS} T b \nonumber
\end{eqnarray}

We can non-dimensionalize these equations by introducing scaled field variables: $T = b^*\tilde{T}$, $b = b^*\tilde{b}$, $S = b^*\tilde{S}$, where $b^*$ is the total density of attachment sites (including both available and attached), assumed to be constant along the hypha. We also scale time and distance: $t = \tilde{t}/k_{ST}$ and $x = \tilde{x}U/k_{ST}$, to arrive at the dimensionless equation:
\begin{eqnarray} 
\frac{\partial \tilde{T}}{\partial \tilde{t}} + \frac{\partial \tilde{T}}{\partial \tilde{x}} &=& -\alpha_1 \tilde{T}\tilde{b} + \tilde{S} \nonumber \\
\frac{\partial \tilde{S}}{\partial \tilde{t}}&=& \alpha_1 \tilde{T}\tilde{b} - \tilde{S} \label{eq:pdedimless} \\
\frac{\partial \tilde{b}}{\partial \tilde{t}} &=& \tilde{S} - \alpha_1 \tilde{T} \tilde{b} \nonumber
\end{eqnarray}

This set of equations have no soliton solutions. Specifically, we look for solutions in which $\tilde{T}(\tilde{x},\tilde{t}) \equiv \tilde{T}(\tilde{x}-c\tilde{t})$, $\tilde{b}(\tilde{x},\tilde{t}) \equiv \tilde{b}(\tilde{x}-c\tilde{t})$ etc. under the constraint that there is a constant $T_\infty$, such that $\tilde{T}\to T_\infty$, as $\tilde{x}\to \pm \infty$: i.e. the far up-stream and down-stream values for $\tilde{T}$ should be identical, similarly for $\tilde{b}$, $\tilde{S}$. Then we can use the fact that under our non-dimensionalization: $\tilde{S}+\tilde{b}=1$ to reduce to two ODEs:
\begin{eqnarray} 
(1-c) \frac{dT}{d\xi} &=& -\alpha_1 T(1-S) + S \nonumber \\
-c\frac{dS}{d\xi}&=& \alpha_1 T(1-S) - S \label{eq:travwave}
\end{eqnarray}
where $\xi = \tilde{x}-c\tilde{t}$, and we use the dimensionless forms of each variable, but suppress the tildes. For there to exist a traveling wave solution the original PDEs, there must exist a solution of these equations for some choice of the parameter $c$. How many boundary conditions are there? To determine the number of boundary conditions we must apply as $\xi\to \pm \infty$, we linearize Eqs.(\ref{eq:travwave}) by writing $T(\xi)= T_\infty + T'(\xi)$ with $|T'(\xi)|\ll T_\infty$, etc, and keep only linear terms, to deduce that $\alpha_1T_\infty(1-S_\infty) = S_\infty$ and:
\begin{equation}
\frac{d}{d\xi}\left(\begin{array}{c} T' \\ S' \end{array}\right) = \left(\begin{array}{cc} -\frac{\alpha_1(1-S_\infty)}{1-c} & \frac{\alpha_1 S_\infty + 1}{1-c} \\ -\frac{\alpha_1(1-S_\infty)}{c} & \frac{\alpha_1 S_\infty +1}{c} \end{array}\right)\left(\begin{array}{c} T' \\ S' \end{array}\right) \label{eq:linearinfty}
\end{equation}
The eigenvalues of the matrix in Eq.(\ref{eq:linearinfty}) are $\lambda_1=0$ and $\lambda_2 = \frac{1+\alpha_1 T_\infty - c(1+\alpha_1(1+T_\infty-S_\infty))}{c(1-c)}$. The zero eigenvalue of the matrix is associated with the fact that multiple values of $(T_\infty,S_\infty)$ are possible. The other eigenvalue represents a solution that is exponentially growing either as $\xi\to\infty$ or $\xi\to-\infty$, according as $\lambda\gtrless 0$, and exponentially decaying in the other direction. Suppose $\lambda_2>0$, then if we were to solve Eqs.(\ref{eq:travwave}) by shooting, starting at $\xi = \infty$, then there are no free parameters to shoot on (no-exponentially decaying linearized solutions). Although $c$ appears to be a free parameter, it does not affect this result, since Eqs.(\ref{eq:travwave}) are also translationally invariant in $\xi$, and so we must have both free choice of $c$, and an exponentially decaying solution in order to generate a solution of the ODEs that is consistent with the boundary conditions.

Although unable to support solitons, the system of equations (\ref{eq:pdedimless}) is consistent with the anti-congestion measured in real fungal hyphae. To compare with our measured data on the dependence of nuclear velocity upon density, we analyze the propagation of long wavelength disturbances (for which the length scale of the disturbance: $\frac{T}{\frac{dT}{dx}}\gg 1$). Then the two populations $T$ and $S$ are in local equilibrium, meaning that $\alpha_1T(1-S) = S$ and if the total number of nuclei (both attached and free-flowing) at some point in the hypha is $n = T+S$, then:
\begin{equation}
u(n) \equiv \frac{T}{n} = -\frac{\alpha_1(1-n)+1+\sqrt{\left(\alpha_1(1-n)+1 \right)^2+4\alpha_1 n}}{2\alpha_1} \label{eq:un}
\end{equation}
and on summing the two equations in (\ref{eq:pdedimless}) we obtain:
\begin{equation}
\frac{\partial n}{\partial t} + \frac{\partial T}{\partial x} \equiv \frac{\partial n}{\partial t} + \frac{\partial}{\partial x}\left(u(n)n \right) = 0 \label{eq:conserv}
\end{equation}
Eq. (\ref{eq:conserv}) is a flux-conservation law with a velocity of transport $u(n)$ that increases with the density of nuclei, $n$. It can also be fit to the real data (Fig. 2B) in the main paper, since it is identical to the velocity-density law discussed in Section \ref{sec:remodel} under the rescaling $n(1+1/\alpha_2) \to n$.

\subsection{Including remodeling of attachment sites \label{sec:remodel}}

We infer from the tendency of nuclei to flow with the cytoplasm for a time interval $\tau$ of almost 20s following the passage of a anti-jam, that attachment sites are not immediately able to accept a new nucleus following detachment of the previous one. We therefore incorporate into our model a population, $u$, of attachment sites with no nucleus currently attached but that are not able to accept a new nucleus, and model the attachment and detachment processes using a set of rate equations:
\ce{$T+b$ ->[k_{TS}] $S$ ->[k_{ST}] $T+u$}, \ce{$u$ ->[1/\tau] $b$}
since $b+S+u=b^*$, under the same non-dimensionalization as in Section \ref{sec:noremodel} we obtain equations:
\begin{eqnarray} 
\frac{\partial T}{\partial t} + \frac{\partial T}{\partial x} &=& -\alpha_1 T b + S \nonumber \\
\frac{\partial S}{\partial t}&=& \alpha_1 Tb - S \label{eq:delaypde} \\
\frac{\partial b}{\partial t} &=& \alpha_2(1-S-b) - \alpha_1T b \nonumber
\end{eqnarray}
where we use the same symbols for dimensionless as for dimensional variables. Remodeling adds another dimensionless group $\alpha_2 \equiv 1/k_{ST}\tau$ representing the ratio of the rate at which attachment sites reconfigure to accept new nuclei to the rate at which nuclei disassociate from attachment sites. We can analyze Eqs.(\ref{eq:delaypde}) in the same long wavelength limit as Eq.(\ref{eq:un}) to derive a flux conservation equation: $\frac{\partial n}{\partial t} + \frac{\partial}{\partial x}\left( u(n)n\right) = 0$ with:
\begin{equation}
u(n) \equiv \frac{T}{n} = -\frac{\left(1+\alpha_1\left(1-n\left(1+\frac{1}{\alpha_2}\right)\right)\right)+\sqrt{\left[1+\alpha_1\left(1-n\left(1+\frac{1}{\alpha_2}\right)\right)\right]^2+4\alpha_1 n\left(1+\frac{1}{\alpha_2}\right)}}{2\alpha_1} \label{eq:undelay}
\end{equation}
$u(n)$ increases with $n$, so the model correctly predicts that nuclei flow anticongestively. We can fit the theoretical curve to real nuclear velocity-density data (Fig. 2B in the main text) by inputting the measured velocity of cytoplasmic flow for this hyphae $U=9.6\,\mu$ms$^{-1}$ and setting $b^*=0.22\,\mu$m$^{-1}$ (i.e. one potential attachment site every 5 $\mu$m). 

We solve the PDEs (\ref{eq:delaypde}) numerically using the method of lines on a regular grid, i.e. write the PDEs as a set of coupled ODEs for $(T,S,b)$ evaluated at a regularly spaced set of points $x_i$. The derivative term $\frac{\partial T}{\partial x}$ is represented by second order upwinded finite differences, and the ODEs are integrated using a fully implicit ODE solver (the Matlab function ode15s).

\section{Nuclear profile in an anti-jam \label{sec:anti-jams}}

As $\alpha_2\to 0$, the solution for the traveling wave equation can be expanded into inner and outer regions, describing nuclear dynamics within the anti-jam itself, and in its wake. The dynamics of these two regions correspond to different dominant balances of the terms in Eqs.[\ref{eq:pde}]. In the anti-jam wake, we need to scale distance by introducing $\Xi=\alpha_2\xi$. To access the dynamics of the two different regions, we expand the variables representing the populations of free-flowing and bound nuclei and attachment sites as: $T = T^{(0)}(\xi) + O(\alpha_2)$, $S = S^{(0)}(\xi) + O(\alpha_2)$, $b = b^{(0)}(\xi) + b^{(1)}(\Xi)+O(\alpha_2)$. The first terms in each of these expansions are given by Eqs. from the main-text, in which transitions $u\to b$ are neglected, so that the density of available attachment sites $b^{(0}(\xi)$ decays monotonically from the front of the anti-jam to its rear.

In the wake of the anti-jam, available attachment sites are regenerated by the $u\to b$ transition according to a dominant balance equation:
\begin{equation}
-c\frac{db^{(1)}}{d\Xi} = 1-b^{(1)}
\end{equation}
so that $b^{(1)}(\Xi) = 1-\exp(\Xi/c)$. 

In Fig. 2A in the maintext we compare the approximate soliton solution (in the limit of $\alpha_2\to 0$) with the measured profiles of 11 real anti-jams. In fitting theory to measured profiles we have two free parameters -- the total number of nuclei in the traveling wave and the dimensional length scale $Uk_{ST}$. We find the best fit if $U/k_{ST} = 70 \mu$m. This is in adequate agreement with the independently measured value of these parameters $U/k_{ST} = 190 \mu$m. The velocity of the anti-jam can be related to the total number of nuclei that it contains, and in principle this gives an additional constraint on the model. However, because of nuclear rearrangements within the anti-jam, and changes in its velocity as it passes through septal pores etc, we were unable to constrain the anti-jam velocity any more precisely than the range $0.8U<c<U$, meaning that the number of nuclei had to remain as a fitting parameter in our model.

\section{Stability analysis of uniform flow \label{sec:stability}}

\begin{figure}
\begin{center}
\includegraphics[width=8cm]{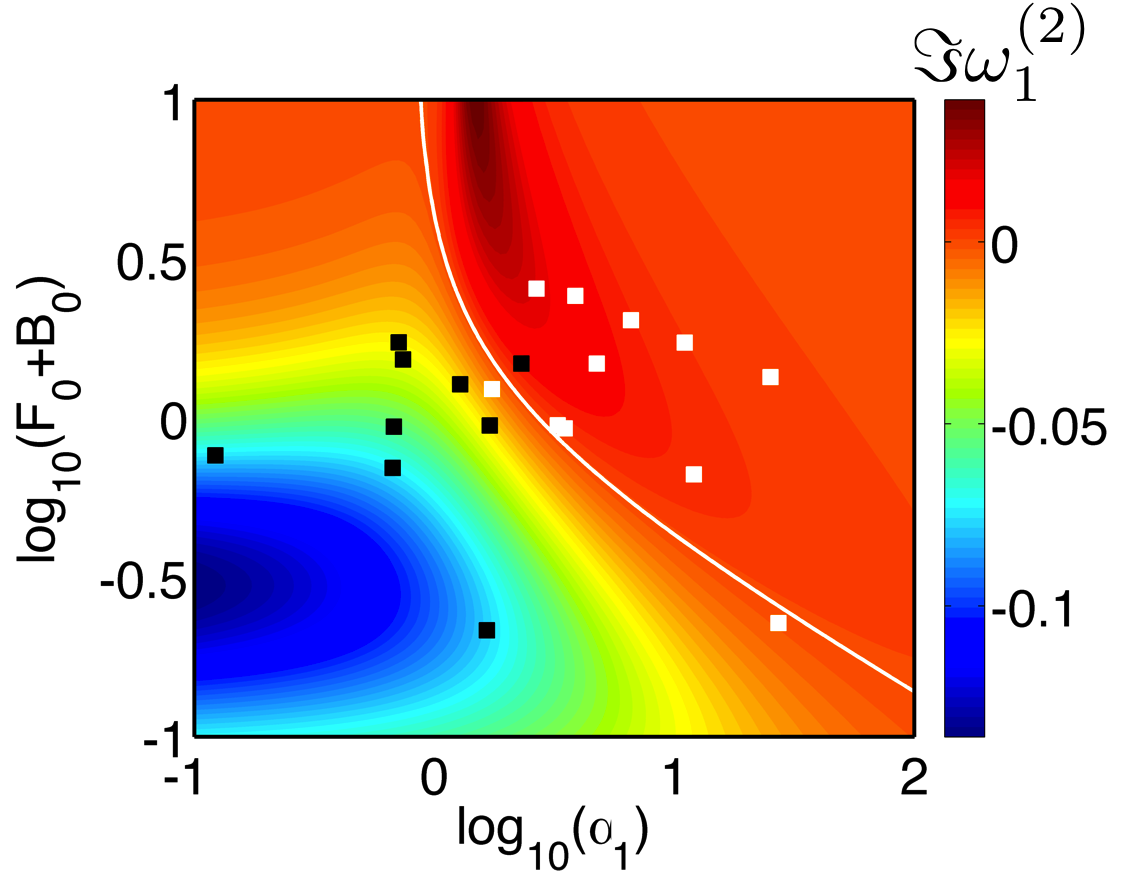}
\caption{Stability boundary and growth rate based on $O(k^2)$ expansion, to be compared with Fig. 4B in the main text. Symbols represent experimental data, as shown in the main-text.} \label{fig:stabbdry}
\end{center}
\end{figure}

In hypha carrying a uniform density $N=T+S$ of nuclei and with uniform density $b+S=1$ of attachment sites, we consider the effect of a small density perturbation within the hyphae. Specifically since the PDEs (\ref{eq:delaypde}) do not contain $x$ or $t$ explicitly, we can decompose the perturbation into monochromatic functions:
$T = T_0 + \tilde{T}e^{i(kx-\omega t)}$, etc., where $T_0$, $S_0$, and $b_0$ are constant solutions of the PDEs (which were already calculated for the long-wavelength analysis carried out in Section \ref{sec:remodel}). Then the perturbation amplitudes $\tilde{T}$, $\tilde{B}$, $\tilde{b}$ satisfy the matrix equation:
\begin{equation}
\left(\begin{array}{ccc} i(k-\omega)+\alpha_1 b_0 & -1 & \alpha_1 T_0 \\ 
-\alpha_1 b_0 & -i\omega+1 & -\alpha_1 T_0 \\
\alpha_1 b_0 & \alpha_2 & -i\omega +\alpha_2+\alpha_1 T_0
\end{array}\right) \left( \begin{array}{c} \tilde{T} \\ \tilde{S} \\ \tilde{b} \end{array} \right) = \bzero \label{eq:evalue}
\end{equation}
This equation has a non-trivial solution only if $\omega$ is an eigenvalue of the matrix. Given any value of $k$ there are three such eigenvalues. One eigenvalue, $\omega_1(k)\to 0$ as $k\to 0$ (representing the existence of multiple constant solutions of the PDEs), while the other eigenvalues $\omega_1(k)$, $\omega_2(k)$ have purely imaginary limits, but with negative imaginary part (i.e. are stable). We find that there is a finite range of $k$-values over which the $\Im(\omega_1(k))$ is positive (shown in Fig. 4A), but that $\Im(\omega_2(k))$, $\Im(\omega_3(k))<0$ for all $k$. Thus $\Im \omega(k)$ is bounded above. In Fig. 4B we plot the numerically determined maximum eigenvalue of (\ref{eq:evalue}), i.e. the maximum growth rate of any perturbation.  To calculate this plot, for each parameter pair ($\alpha_1,\,T_0+S_0$) we perform a sweep over values of $k$ to find the unstable wave numbers, if any, and locate the maximum. However, the most important feature of this plot is the stability boundary -- i.e. the curve separating stable parameter regimes ($\max \Im \omega=0$) from unstable parameter regimes (which $\max \Im \omega>0$). We can calculate this boundary without any parameter sweep, by determining whether $\Im \omega_1(k) \to 0\pm$ as $k\to 0+$ (i.e. where the curve is initially decreasing or increasing), using a perturbation analysis of (\ref{eq:evalue}) around $k=0$. Namely we expand $\omega_1(k)$ as a power series: $\omega_1(k) = k \omega_1^{(1)} + k^2 \omega_1^{(2)}$. If $\bv_1 = \bv_1^{(0)} + k\bv_1^{(1)}+ k^2\bv_1^{(2)}$ is the corresponding eigenvector and $\bv_2^{(0)}$, $\bv_3^{(0)}$ are the other eigenvectors for $k=0$, and we define matrices:
\begin{equation}
A = \left(\begin{array}{ccc} -i\alpha_1 b_0 & i & -i\alpha_1 T_0 \\ 
i\alpha_1 b_0 & -i & i\alpha_1 T_0 \\
-i\alpha_1 b_0 & -i\alpha_2 & -i\alpha_2-i\alpha_1 T_0
\end{array}\right) \quad \hbox{and} \quad D = \left(\begin{array}{ccc} 1 & 0 & 0 \\ 0 & 0 & 0 \\ 0 & 0 & 0 \end{array} \right)~,
\end{equation}
then (\ref{eq:evalue}) can be written as:
\begin{equation}
(A + kD) (\bv_1^{(0)}+k\bv_1^{(1)}+k^2\bv_1^{(2)}) = (k \omega_1^{(1)} + k^2 \omega_1^{(2)})(\bv_1^{(0)}+k\bv_1^{(1)}+k^2\bv_1^{(2)})~.
\end{equation}
We collect together like powers of $k$, obtaining at $O(k)$:
\begin{equation}
D\bv_1^{(0)}  =  \omega_1^{(1)} \bv_1^{(0)} - A \bv_1^{(1)} 
\end{equation}
since $A$ is invertible in the subspace $V$ $\bv_2^{(0)}$, $\bv_3^{(0)}$, this equation can be solved for $\omega_1^{(1)}$ and $\bv_1^{(1)}$, if we assume that $\bv_1^{(1)}\in V$, which is permissible since any component of $\bv_1^{(1)}$ outside of $V$ can be absorbed into $\bv_1^{(0)}$. Because $A$ is purely imaginary and $D$ and $\bv_1^{(0)}$ are real, $\omega_1^{(1)}$ will be real i.e. corresponds to a traveling wave that neither grows nor decays. Therefore to determine stability we must continue to calculate the$O(k^2)$ terms in the expansion:
\begin{equation}
(D-\omega_1^{(1)})\bv_1^{(1)} =  \omega_1^{(2)} \bv_1^{(0)} - A \bv_1^{(2)} \label{eq:k2}
\end{equation}
since $A$ is invertible on $V$ we can solve (\ref{eq:k2}) for $\omega_1^{(2)}$ and $\bv_1^{(2)}$. The sign of $\Im(\omega_1^{(2)})$ can used to reconstruct the stability boundary of the uniform flow state more rapidly and more accurately than performing a $k$-sweep (see Fig. \ref{fig:stabbdry}). To compare the predicted stability boundaries with conditions in which anti-jams are observed in real hypha, we need to infer the values of $\alpha_1$, $T_0+S_0$ for real hyphae (we assume that $\alpha_2=1.25$ for all hypha, based off the measurements of $k_{ST}$ and $\tau$ given in the main text). For our reference hypha: $\alpha_1\equiv k_{TS} b^*/k_{ST} = 3.5$, from the ratio of the rates at which nuclei transition from free-flowing to non-flowing, and non-flowing to free-flowing (Fig. 2D in the main text). The `off-rate' $k_{ST}$ depends only on the thermodynamics of nuclear attachment to the cortical microtubules, and should be the same within and between hyphae, so $\alpha_1$ varies between hyphae only through the `on-rate' $k_{TS} b^*$. The on-rate has a kinetic component -- faster moving nuclei will encounter more potential attachment sites and are more likely to bind. A simple kinetic model, based on hard sphere interactions, gives $k_{TS} b^*\propto $ (hyphal velocity) and $k_{TS} b^*\propto $ (hyphal diameter)$^{-1}$. So for a hypha with velocity $V$ and diameter $d$, we scale $\alpha_1$ from the dimensions of our reference hypha:
\begin{equation}
\alpha_1 = 3.5 \left(\frac{V}{9.81\, \mu{\rm ms}^{-1}}\right) \left(\frac{12.73\, \mu{\rm m}}{d} \right)~.
\end{equation}
The stability boundary also depends upon the dimensionless density of nuclei in the hyphae: $S_{0}+T_{0}$. We directly measure the density of nuclei, by counting the number of nuclei in a 100\,$\mu$m length of hypha. The density fluctuates depending on whether an anti-jam is contained within the 100 $\mu$m reference section or not, and also varies with time depending on the spacing of anti-jams within the hypha (anti-jams do not arrive periodically, see Fig 1D). To get a stable read of nuclear density, we include only measurements in which the reference section contains an anti-jam. Since anti-jams are typically less than 50\,$\mu$m long, the density measurement includes both the anti-jam and the regions ahead and behind it. We non-dimensionalize the nuclear density by the total linear density of potential attachment sites, $b^*$. For the reference hyphae we infer by fitting the long wavelength theory (\ref{eq:undelay}) to the experimental data that $b^*=0.25\, \mu$m. However, assuming that the area of cell membrane per attachment sites is the same for each hyphae, $b^*$ will scale proportional to the surface area-length ratio of each hypha, i.e. $b^*\propto$ (hyphal diameter). Thus:
\begin{equation}
T_{0}+S_{0} = \frac{\hbox{measured density of nuclei}}{0.22\, \mu{\rm m}^{-1}}  ~ \times \left(\frac{12.73\, \mu{\rm m}}{d} \right)
\end{equation}

\begin{figure}
\begin{center}
\includegraphics[width=8cm]{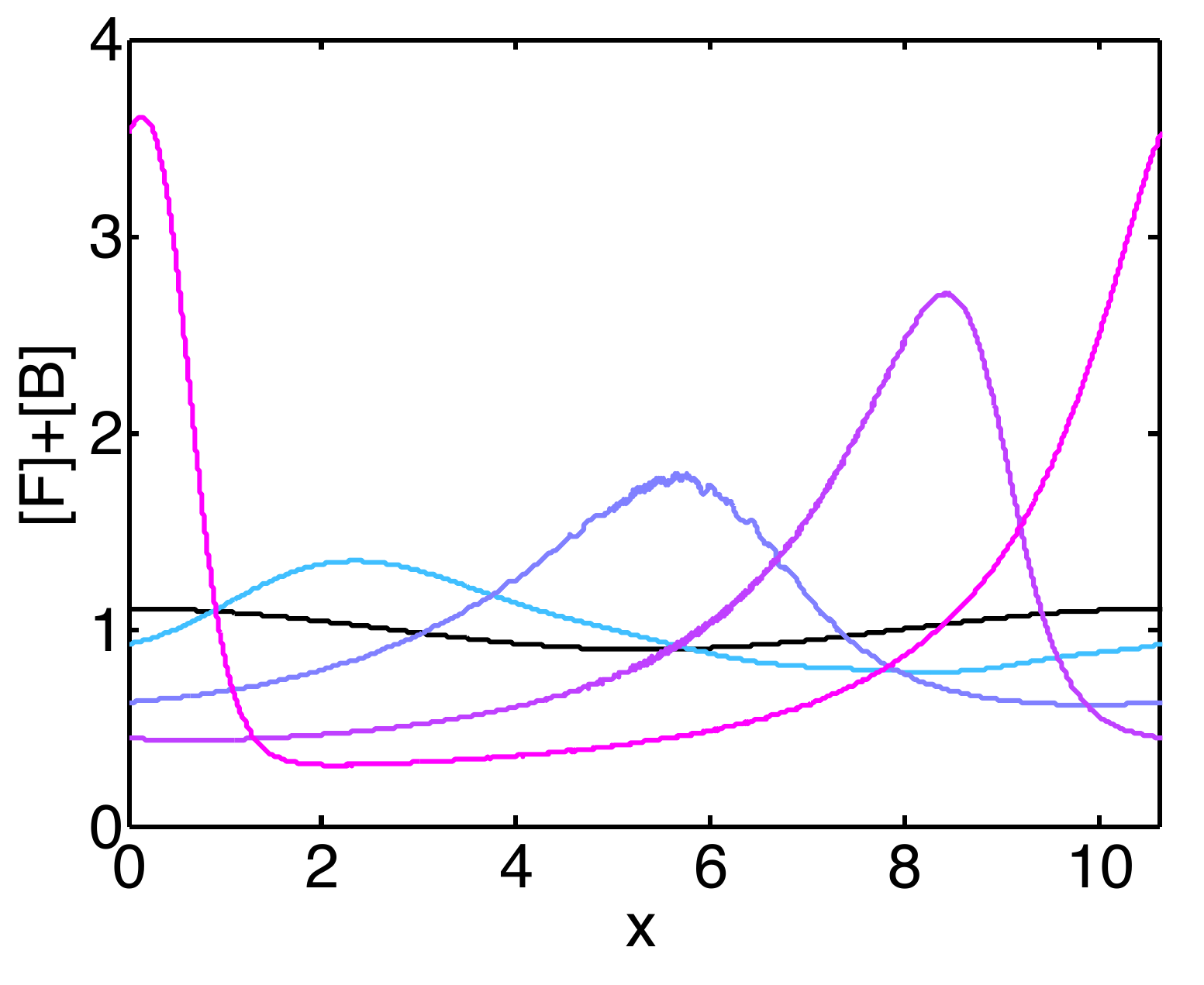}
\caption{Nonlinear growth of the maximally unstable perturbation for parameters $\alpha_1=1$, $\alpha_2=0.1$ and $S+T=1$. The black curve gives the initial density of flowing nuclei, slightly perturbed from uniform, colored curves show numerically simulated evolution toward a soliton/anti-jam at times $t=80$ (blue), 160, 240, 320 (magenta).} \label{fig:nonlinearevoln}
\end{center}
\end{figure}

To study the growth and evolution of perturbations outside of the linear regime, we simulate the fully non-linear PDEs using the method described in Section \ref{sec:remodel}, using the monochromatic perturbation with fastest linear growth rate as an initial condition. If the fastest growing linear perturbation has wavenumber $k_{\rm max}$ we solve the evolution equations in a periodic domain of size $2\pi/k_{\rm max}$. We find that, starting with very small perturbations, the density profile evolves toward a soliton (Fig. \ref{fig:nonlinearevoln}). The soliton solution is described mathematically in the main-text and in Section \ref{sec:anti-jams}.
